# Photoluminescence properties of Er-doped AlN films prepared by magnetron sputtering

H. Rinnert[(*)], S. S. Hussain, V. Brien, and P. Pigeat

Institut Jean Lamour, Nancy-University, UPVM, CNRS, Boulevard des Aiguillettes

B.P. 239, 54506 Vandœuvre-lès-Nancy Cedex, France.

## Abstract

Er-doped aluminum nitride films, containing different Er concentrations, were obtained at room temperature by reactive radio frequency magnetron sputtering. The prepared samples show a nano-columnar microstructure and the size of the columns is dependent on the magnetron power. The Er-related photoluminescence (PL) was studied in relation with the temperature, the Er content and the microstructure. Steady-state PL, PL excitation spectroscopy and time-resolved PL were performed. Both visible and near infrared PL were obtained at room temperature for the as-deposited samples. It is demonstrated that the PL intensity reaches a maximum for an Er concentration equal to 1 at. % and that the PL efficiency is an increasing function of the magnetron power. Decay time measurements show the important role of defect related non radiative recombination, assumed to be correlated to the presence of grain boundaries. Moreover PL excitation results demonstrate that an indirect excitation of $Er^{3+}$ ions occurs for excitation wavelengths lower than 600 nm. It is also suggested that Er ions occupy at least two different sites in the AlN host matrix.

---

[*] Corresponding author : herve.rinnert@ijl.nancy-universite.fr

**PACS Code:** 78.55.-m / 78.47.jd / 81.15.Cd / 77.55.hd

**Keywords**: Aluminium nitride, erbium doping, R. F. sputtering, photoluminescence.

**I. Introduction**

Over the last decade rare-earth (RE)-doped semiconductors have been intensively studied because of their potential applications in opto-electronics. Some applications require a low-temperature elaboration process. Rare earth elements have partially filled 4$f$ shells which are screened by 5$s$ and 5$p$ electronic states. Due to their shielded 4$f$ levels, rare earth ions emission is consequently insensitive to the host matrix. Among the RE elements, Er is of prime interest because of its $^4I_{13/2} \rightarrow {}^4I_{15/2}$ electronic transition that gives rise to an emission at 1.5 µm, which coincides with the main low-loss region in the absorption spectrum of silica-based optical fibres. Moreover $Er^{3+}$ ions can also produce light in the visible range at 558 nm (green, one of the primary colours) which render this ion interesting for electroluminescent devices for use in display technologies.

However, the thermal quenching strongly reduces the radiative emission at room temperature, inducing a very weak Er related emission in bulk silicon for example. It was found that the thermal quenching effect in RE-doped semiconductors is related to the band gap energy value of the material.[1] As the temperature quenching is weak for wide gap materials, doping AlN with rare earth elements is promising for applications. Moreover wide gap materials also render possible light emission in the visible range.

The luminescence of Er ions inside an AlN matrix has firstly been demonstrated in epitaxially grown AlN, in which Er was introduced by implantation.[2] Er-doped AlN thin films were also obtained by metal organic molecular beam epitaxy[3,4], in which an indirect excitation of $Er^{3+}$ ions was demonstrated. R.F. magnetron sputtering was also used to obtain Er doped AlN layers.[5,6,7] In this case samples are amorphous and the Er related PL is weak or absent. An annealing treatment above 900 °C generally leads to the characteristic emission lines of $Er^{3+}$ ions. This emission can be explained by the decrease of the non radiative defect number and to the optical activation of $Er^{3+}$ ions. More recently, by depositing on substrate maintained at 300 °C, Liu et al. have obtained crystalline AlN:Er films by R. F. sputtering.[8]

In this paper we show that it is possible to obtain an intense Er-related PL from as-deposited samples prepared at room temperature by R. F. magnetron sputtering, provided the magnetron power is correctly chosen. The Er content was precisely modified using different targets. It is demonstrated that the PL intensity reaches a maximum for an Er concentration equal to 1 at. %. The obtained samples show a nanocolumnar morphology with the würtzite structure. It is demonstrated that an indirect excitation process of Er ions occurs and that the non radiative channels mainly govern the PL emission. The existence of two different Er sites is proposed to explain the film microstructure dependence of the PL properties. The PL is characterized by steady state PL, time resolved PL, temperature dependent PL and PL excitation experiments.

**II. Experiments**

AlN:Er films were deposited using a R. F. magnetron sputtering at room temperature on Si (100) substrates in a gas mixture of Ar and $N_2$. The $N_2/(Ar+N_2)$ percentage in the gas mixture was set to 50 % and the total pressure is equal to 0.5 Pa. The target-sample distance is equal to 150 mm. Before deposition, the target was sputter cleaned for 15 min using an Ar plasma. The control of the thickness is achieved by real time optical interferometry. Samples thickness is equal to 500 nm. In this study the magnetron power $P_m$ was varied from 50 W to 600 W. No thermal annealing treatment was performed after deposition. Er was introduced by co-deposition using different specific targets containing Er sectors with adequate surface. Each specific target is then associated to an Er content in the film.

The Er concentration was determined by Electron Dispersive Spectroscopy of X-rays (EDSX) and by the highly sensitive Rutherford backscattering spectrometry (RBS) technique. The Er content was varied from 0.5 to 3.6 at.%. The crystalline structure is observed by transmission electron microscopy (TEM). TEM observations were performed on a CM20 PHILIPS microscope operating at an accelerating voltage of 200 kV. For the steady state PL experiments, two different setups were used depending on the wavelength range. For the PL in the visible range, the samples were excited by the UV lines (314 and 334 nm) of a mercury arc lamp source and the detection system was equipped with a charge coupled camera cooled at 140 K. For the infrared PL, the samples were excited by a 30 mW He-Cd laser using the 325 nm line. The PL signal was analyzed by a monochromator equipped with a 600 grooves/mm grating and by a photomultiplier tube cooled at 190 K. For the PL excitation (PLE) experiments, the samples were excited by an optical parametric oscillator (OPO) laser. For the time-resolved PL experiments, the samples were pumped by the 355 nm line of a frequency-tripled YAG:Nd laser. The laser pulse frequency, energy and duration were typically equal to 10 Hz, 50 µJ and 20 ns, respectively. The time response of the detection system was better than 0.1 µs. The spectral response of the detection system was precisely calibrated with a tungsten wire calibration source.

**III. Results and discussion**

Figure 1 (a) and (b) show cross sections views obtained by TEM for samples prepared with $P_m$=50 W and $P_m$=600 W, respectively. The micrographs show the columnar structure of the AlN layers. It is rather difficult to discern an obvious evolution between the different microstructures, but the average width of the columns is a progressive and increasing function of the magnetron power. The size of the columns, measured at the top of the layers, is around 40 nm and 60 nm for $P_m$=50 W and $P_m$=300 W, respectively.

The RE content is known to strongly influence the PL properties. Hence, the Er concentration was changed to determine the optimal preparation condition. Figure 2 (a) and (b) show the dependence of the Er content on the visible and near infrared PL for samples prepared with $P_m$=300 W. The spectra show PL emission at 537, 558, 667, and around 1.5 µm, corresponding to the radiative electronic transitions of $Er^{3+}$ ions from the $^2H_{11/2}$, $^4S_{3/2}$, $^4F_{9/2}$ and $^4I_{13/2}$ states to the $^4I_{15/2}$ ground state, respectively. The inset of Fig. 2 (b) shows the PL intensity at 1.53 µm as a function of the Er concentration and demonstrates the existence of an optimum concentration at around 1 at.%. For lower concentration, the Er PL intensity is proportional to the Er concentration which is due to the increase of the emission centres. For higher content the decrease of the PL can be explained by non radiative interactions between close neighbor $Er^{3+}$ ions, which quench the luminescence. From this point, all PL results that are presented thereafter correspond to Er-doped samples prepared with a concentration equal to 1 at.%.

Figure 3 (a) shows the evolution of the spectra in the range 77 K-300 K. Figure 3 (b) represents the temperature dependence of the PL intensity at the peak maximum located at 1.53 µm and the integrated PL in the range 1450-1625 nm. The results evidence a decrease of the PL intensity at 1.53 µm by a factor of 2 from 77 K to room temperature. However the evolution of the integrated PL shows a decrease by a factor of only 1.15 in the same temperature range. Note that for comparison purposes, the scaling factor is the same for the left and right axis. This behaviour is explained by the fact that the peaks located at 1.51 µm and at 1.53 µm follow a different trend as shown in Fig.3 (a). The high energy part of the $^4I_{13/2}$ → $^4I_{15/2}$ transition is an increasing function of the temperature which is due to the temperature dependence of the occupation probability of the states. The total disappearance of the high energy contribution of the PL band has already been mentioned at 10 K.[7] The negligible loss of the total PL intensity obtained here from 300 K to 77 K is in agreement with the predicted weak temperature quenching for wide gap materials.[1]

The PLE spectrum is shown in Fig. 4 for a sample prepared with $P_m$ = 600W. For this experiment the PL was measured at 1.53 µm. The sharp structures shown here suggest that $Er^{3+}$ ions can be excited directly. As observed at 15 K in a previous study[4], the bands at 462, 499, 535, and 676 nm are attributed to the excitation of $Er^{3+}$ ions from the ground state to the $^4F_{3/2}$ and $^4F_{5/2}$, $^4F_{7/2}$, $^2H_{11/2}$ and $^4F_{9/2}$, respectively. The other contribution at 558 nm observed here can be attributed to the excitation from the ground state to the $^4S_{3/2}$ state. Moreover the inset of Fig. 4 clearly shows that these bands present two well defined contributions. A broad band is also observed from around 600 nm to shorter wavelengths, which shows the occurrence, at room temperature, of a non resonant excitation of $Er^{3+}$ ions, demonstrating an indirect excitation of the ions. This indirect process is a decreasing function of the excitation wavelength. As proposed in Er-doped GaAs films,[9] this indirect excitation could be due to carrier-mediated processes.

Figure 5 (a) and (b) show the dependence of the magnetron power on the PL emission in the visible and near infrared range, respectively. A broad PL band is obtained between 450 and 500 nm. The intensity of this contribution is a decreasing function of $P_m$. The integrated intensity of this band decreases by more than two orders of magnitude from $P_m$=50 W to $P_m$=600 W. This band was already mentioned in previous studies but there is no consensus on the physical origin of these radiative transitions. It has been attributed to the existence of native defects, to nitrogen vacancies, to the presence of oxygen or to interface defects.[10,11,12] In our case, the variation of PL intensity of this band can not be attributed to oxygen because the chemical composition shows no significant change in oxygen content from one sample to another. This band may be interpreted by the existence of transitions between electronic states located in the bandgap of the AlN matrix. It is well known that an amorphous or disordered structure involves the existence of band tails states which introduce electronic states inside the bandgap. The structural study suggests that the crystalline volume fraction is an increasing function of $P_m$, which justifies the attribution of this band to structural defects associated with the disorder. This band is not well defined, since its exact position and shape of the contributions vary from sample to sample. Moreover, the luminescence spectrum of the film prepared with $P_m$=50 W shows a shoulder around 620 nm. Given the thickness of the films, which is approximately equal to 500 nm, an interference phenomenon is certainly responsible for the modulation and the different shapes of the broad PL band.

The Er-related PL is also dependent on the magnetron power, in agreement with already published results[8]. For the film prepared with $P_m$ = 50 W, the $Er^{3+}$ ions contribution in the visible range is missing or presents an intensity below the detection limit of the measuring setup. The PL intensity at 1.53 micron is very weak. For higher $P_m$ values, the Er-related PL is an increasing function of $P_m$. This increase can be correlated to the change in the column size and to the induced increase of the crystalline volume fraction.

Decay time experiments were performed to interpret this behavior. Figure 6 shows the time-dependence of the PL at 1.53 μm for samples prepared with $P_m$ = 50 W and $P_m$ = 600 W, respectively. The results can not be correctly fitted by only one exponential curve. However, our simulations show that a good agreement is obtained taking into account of two exponential curves. The time dependence of the PL intensity can be written as follows:

$I_{PL}=I_f \exp(-t/\tau_f) + I_s \exp(-t/\tau_s)$,

where *f* and *s* characterize the fast and slow PL components. The curve fits are represented in the graph.

These results show that the characteristic PL decay time value is in the range of tenths of microseconds for sample prepared at low power. A strong increase of the decay time is obtained by increasing the magnetron power. As the decay time increase is correlated to the PL intensity increase, the PL is governed by non radiative recombination channels. The capture radius of the defects, probably located in the grain boundaries, is not known but it can be assumed that if $Er^{3+}$ ions are located inside the well crystallized columns, far from the grain boundaries, the efficiency of the defects is weak. Assuming that $Er^{3+}$ ions are uniformly distributed in the structure, the column size increase induces an increasing number of $Er^{3+}$ ions located inside the columns and then at a larger distance from the grain boundaries. Consequently the PL decay time and the PL intensity are increasing functions of $P_m$.

The spectroscopic results suggest that the samples can be simply modeled by considering two different $Er^{3+}$ sites. It can be assumed that erbium ions could be localized in substitution of the Al atoms inside the würtzite, or in insertion site in either the octahedral site or the tetrahedral one. The existence of two sites involves two different local environments and crystal field effects for the ions, which explain some difference in the optical activation of the rare earth ions and to the existence of two different radiative decay times. The measured decay time increase from 6 to 39 μs and from 66 to 163 μs for the samples prepared with $P_m =$ 50 W and $P_m =$ 600 W, respectively. These measured values are influenced by the non radiative processes and the radiative decay time could be measured in monocrystalline samples. A non exponential behavior of the PL decay time was already mentioned in epitaxially grown AlN:$Er^{3+}$ samples.[3] In these samples, two decay times have been estimated. The values are equal to 50 and 830 μs and are then in agreement with our values. The existence of two different sites, where different stark splitting effects occur, should induce different spectral signature. Further experiments at low temperatures will be performed to investigate the influence of the crystal field effect. Moreover, in the PLE spectrum, the lines corresponding to the direct excitation of $Er^{3+}$ ions are relatively broad. This result is in good agreement with the assumption of the existence of two different sites for the localization of $Er^{3+}$ ions.

## IV. Summary

Er-doped AlN thin films were obtained at room temperature by reactive radio frequency magnetron sputtering. A strong Er-related PL is obtained at room temperature for all as-deposited samples studied. The PL is characterized by a very weak temperature quenching, which is probably due to the wide gap of the AlN matrix. The PL intensity was found to be optimized with an Er concentration equal to 1 at.%. Above this value a concentration quenching occurs and limits the PL emission intensity. The near infrared Er-related PL emission is an increasing function of the magnetron power. This result is tentatively explained by the increase of the column size of the AlN matrix, favoured by high magnetron power values. In these samples, the grain boundaries probably act as non radiative centres, which play a major role in the PL emission process. For large column sizes, $Er^{3+}$ ions are mainly located in the crystalline AlN columns and the efficiency of non radiative processes is weak. It is also demonstrated that $Er^{3+}$ can be excited either directly by intra $4f$ electronic transitions from the ground state to excited states or by indirect excitation processes. The latter could be due to carrier-mediated energy transfer processes. Finally, both PL excitation and PL decay time experiments suggest that $Er^{3+}$ ions occupy at least two different sites. Further experiments are needed to investigate the influence of the crystal field effects, which are dependent on the $Er^{3+}$ localization in the matrix.

**Figure captions**

Figure 1 : Dark field cross-section views obtained by TEM for (a) sample prepared at $P_m$=50 W and (b) samples prepared at $P_m$=600 W.

Figure 2 : Influence of the Er concentration on the PL properties, (a) in the visible range and (b) around 1.5 µm.

Figure 3 : Temperature dependence of the PL of sample prepared with $P_m$=300 W and containing 1 at.% Er.

Figure 4 : Photoluminescence excitation spectrum measured at 1.53 µm for sample prepared with $P_m$=600 W and containing 1 at.% Er.

Figure 5 : Influence of the he magnetron power on the PL emission (a) in the visible range and (b) at 1.5 µm.

Figure 6 : Time-dependence of the PL at 1.53 µm for samples prepared at $P_m$=50 W and $P_m$=600 W.